\documentclass[%
preprint,
 amsmath,amssymb,
 aps,
]{revtex4-2}

\usepackage{graphicx}
\usepackage{ulem}

\usepackage{xcolor}
\newcommand{\rf}[1]{(\ref{#1})}

\newcommand{\beq}{\begin{equation}}
\newcommand{\eeq}{\end{equation}}
\newcommand{\beqr}{\begin{eqnarray}}
\newcommand{\eeqr}{\end{eqnarray}}
\newcommand{\lb}[1]{\label{#1}}
\newcommand{\bc}{\begin{center}}
\newcommand{\ec}{\end{center}}
\newcommand{\ct}[1]{\cite{#1}}

\begin{document}

\title{ Quantum analysis of a multi-frequency laser with a photonic time crystal }

%\maketitle

\author{Igor E. Protsenko}
\email{procenkoie@lebedev.ru}

\author{Alexander V. Uskov}

\affiliation{%
P.N.Lebedev Physical Institute of the RAS,
Moscow 119991, Russia
}%

\begin{abstract}
The present study considers the operation of a laser that incorporates a photonic time crystal (PTC), the purpose of which is to generate a field characterised by multiple widely separated optical frequencies. This laser is the subject of both a proposal and theoretical investigation. The laser comprises an active medium and a PTC within a small cavity constructed from two photonic crystals that are positioned in an overlapping configuration. PTC is modulated by an external field. The spikes in the laser field spectrum are separated by the PTC modulation frequency. The development of a quantum model of the laser with PTC has been achieved, and the analysis of a lasing mode with multi-frequency spikes has been made. The investigation focused on the study of lasing conditions, output power, and the lasing field spectra. The experimental realization of the multi-frequency laser with PTC under realistic conditions is discussed.  
\begin{description}
\item[Keywords] Photonic time crystal, nanolasers, quantum noise
\end{description}

\end{abstract}

\maketitle

\section{\label{Sec1}Introduction}
A photonic time crystal (PTC) \ct{10.1117/1.AP.4.1.014002}
is a time-varying medium \ct{10121054} in which the electromagnetic permittivity $\varepsilon$ or/and permeability $\mu$ are periodically modulated in time, rather than in space, as in conventional spatial photonic crystals \ct{PhysRevLett.58.2059,PhysRevLett.58.2486}. PTC has created a temporal analogue of a spatial photonic crystal, leading to novel wave phenomena and potential applications in optics and photonics. 
The refractive index variation is the cause of momentum bandgaps in PTC, where specific regions (bands) of the wave number $k$ values are prohibited \ct{Asgari:24,Lustig:23,Boltasseva:24,annurev:/content/journals/10.1146/annurev-conmatphys-040721-015537}.  
In a manner analogous to the manner in which spatial crystals disrupt the continuity of space symmetry, PTCs disrupt the continuity of time symmetry, thereby giving rise to novel conservation laws as Floquet energy quasi-conservation \ct{annurev:/content/journals/10.1146/annurev-conmatphys-040721-015537}.  PTC have demonstrated a number of exotic wave effects, including time-refraction, whereby light is reflected or refracted in time rather than space \ct{Lustig:18}. Furthermore, Floquet-engineered light-matter interactions \ct{PerezGonzalez2025lightmatter} and PTC parametric resonances \ct{PhysRevA.109.063517,doi:10.1021/acsphotonics.4c00607} have been observed, with energy exchange occurring between light waves. The functionality of PTCs has been demonstrated in experimental studies employing rapidly modulated metasurfaces  \ct{doi:10.1126/sciadv.adg7541}. Optically pumped epsilon-near-zero (ENZ) materials have also been used to demonstrate PTC functionality  \ct{10.1117/1.AP.4.1.014002}. The paper \ct{Wang2025} proposes the experimental realisation of PTC using artificial materials with high-quality resonances. The potential applications of PTC encompass temporal control of light for optical computing and signal processing. This extends to the time domain through metamaterial concepts \ct{10.1117/1.AP.4.1.014002}, quantum light sources \ct{Sustaeta-Osuna2025}, novel light emitters \ct{doi:10.1073/pnas.2119705119} and lasers \ct{Asgari:24,doi:10.1126/science.abo3324,Th_PTC_las}.

The exponential amplification of light is a promising feature of PTC. The emission of light from a radiation source placed inside a PTC has been the subject of study in \ct{doi:10.1126/science.abo3324}, wherein the non-resonant tunable PTC laser was proposed, with exponentially amplified radiation and the linewidth narrowed over time, becoming monochromatic in the middle of the bandgap. Thresholdless laser operation in PTC combined with four-level active medium has been suggested in \ct{Th_PTC_las}. The presence of extremely narrow bands has been predicted in Moiré PTC, with two binary modulations on the refractive index with different modulation periods \ct{4lqd-z567}. Lasing in time-modulated PTC-like media, as a thin impedance layer with the temporal modulation, is considered in \ct{PhysRevApplied.23.024040}.  Consequently, PTC is a promising medium for applications in lasers, though studies of lasing with PTC are still in the early stages. Several laser schemes with PTC are conceivable \ct{Asgari:24}. Further theoretical investigations into various lasers with PTC could help to plan and carry out experiments.

This paper proposes a novel laser incorporating a PTC. To the best of our knowledge, this type of laser with a PTC has not been considered before, and we believe that it can be realized in an experiment in a relatively easy way.  
The laser is based on well-known photonic crystal cavities \ct{nano11113030} with a nonlinear  
medium and an active medium of two-level emitters in the form of quantum dots \ct{5771527}.  We consider the PTC laser scheme to be similar to the well-known optical parametric amplifier containing a nonlinear medium in its cavity \ct{Nagele:20,PhysRevA.50.1627}.

  {In general, a PTC included laser cavity contains a spatially-uniform medium with the periodically-modulated dielectric susceptibility $\chi(t)$. The $\chi(t)$ modulation can be achieved by various ways \ct{PerezGonzalez2025lightmatter,PhysRevA.109.063517, doi:10.1021/acsphotonics.4c00607,doi:10.1126/sciadv.adg7541, Wang2025}. Here we consider the $\chi(t)$ modulation by the nonlinear medium. Such a way of modulation corresponds to realistic experimental conditions, as we will see, and lets us easily estimate PTC laser parameters necessary for the experiment. In principle, other ways of $\chi(t)$ modulation can be considered as in this paper. The  $\chi(t)$-modulation by nonlinear medium makes our model similar with a standard three-mode intra-cavity sum- or difference frequency generation (SFG/DFG) \ct{boyd2020nonlinear}. We compare our PTC laser with the intra-cavity SFG/DFG in Section \ref{subec2_C}.}

The design of the PTC laser is shown in Figure~\ref{Fig1} and explained in Section 2. We derive Maxwell-Bloch equations for such a laser with PTC. The size of a small cavity of the laser is of the order of the optical wavelength. Cavities of this kind are well-developed and widely used in photonic crystal lasers and related devices \ct{BUTT2021107265}. Such a small cavity provides a high electromagnetic field density inside and, therefore, a large nonlinear refractive index in the nonlinear medium of the PTC.  
As the number of photons in the cavity is small, we will formulate the quantum Maxwell-Bloch equations for the PTC laser following \ct{PhysRevA.59.1667, Andre:19,Protsenko_2021,PhysRevA.105.053713}. In this paper we restrict ourselves by the case of relatively small pump of the active medium, when the population inversion is not achieved and the laser works in the LED   {(Light-Emitting Diode)} regime below the semiclassical lasing  threshold. In such case we apply the approach of \ct{Andre:19}, where the fluctuations of populations in the active medium are neglected. The case of a large active medium pump  will be considered in the future, with the help of the approach of \ct{Protsenko_2021}.  In Section 3, we derive the conditions for resonant generation and determine the field modes of a PTC laser operating at several resonant carrier frequencies.   {By the "field mode" or the "lasing mode" we understand a certain spatial configuration of the field in the laser cavity, which is, mathematically, the eigenfunction of the Laplace operator in Maxwell equation and satisfies the cavity boundary conditions on the cavity borders (semi-transparent mirrors), as it is common in laser physics \ct{SLW}. We do not specify precisely the spatial configuration of the field in the mode assuming the mode, with sufficiently high quality factor, exists in the cavity considered.} The presence of many resonant frequencies in the mode is an important feature of our PTC laser, setting it apart from conventional lasers where each lasing mode usually corresponds to a single carrier frequency. In typical well-known multi-frequency lasing, the separation between frequency components is usually much smaller than the  
carrier frequency, 
\ct{LITTLER1992523}. Various techniques can be employed to achieve multi-frequency lasing with relatively small frequency separation, including active frequency shifting \ct{LITTLER1992523}, Raman scattering \ct{VODCHITS2007467} and modulation of semiconductor lasers \ct{CHEN2023129481}.
Unlike well-known multi-frequency lasers with a small frequency separation, the proposed laser with PTC has a large separation between the lasing field frequency components. In our case, the frequency separation is on the same order of magnitude as the optical or near-infrared modulation frequency of the PTC.   {Such a large frequency separation can be met, usually, in nonlinear optical systems, including SFG, DFG, harmonic generation, Raman lasers, and nondegenerate optical parametric oscillators, but not in the conventional longitudinal-mode or low-frequency-modulated lasers. Here we restrict ourselves by the description of a large frequency separation in the longitudinal mode of the PTC laser.} 

A laser with PTC and multiple widely separated generation frequencies will find many applications. For example, applications of matrix-assisted laser desorption ionisation time-of-flight mass spectrometry (MALDI-TOF MS) provide multiple wavelengths for ionisation, leading to better signal-to-noise ratios than single-frequency lasers \ct{Calderaro2014}. The proposed PTC laser can be useful in: differential absorption lidar for gas sensing \ct{Yu:21}, sensitive to different gas molecules; in wavelength-division multiplexing for optical communications \ct{Geng:22}, creating multiple channels for data transmission in fiber optic networks and increasing bandwidth and data capacity; and in other applications of laser sensing, measurements, nonlinear optics, and material processing. 

In the following sections, we will describe the scheme of the laser with PTC, formulate its quantum-mechanical mathematical model, calculate the stationary mean photon numbers, and estimate the parameters of a PTC laser made using readily available materials. We analyse the 
lasing mode wave numbers found with and without a nonlinear medium, as well as the energies of the fields and the field spectra for each carrier frequency in each mode. We then compare the laser fields of different carrier frequencies found at various parameter values. We demonstrate that population inversion is not achieved under the practical conditions considered in this paper, meaning that the PTC laser operates as a light-emitting diode with a few photons in the cavity and a large amount of spontaneous emission in the lasing mode. {So, strictly speaking, our device is a PTC LED rather than a PTC laser. However, the  positive population inversion, necessary for true lasing, is just a matter of the pump increase and does not require any change in the device scheme and in the lasing mode. In the meanwhile, the analysis of the lasing at high pump in quantum regime is more difficult, than in the present paper, it requires considering population fluctuations \ct{Protsenko_2021, PhysRevA.105.053713} neglected here at relatively low pump. To reduce the size of the paper, we will carry out such analysis in the future, using results for PTC lasing mode  found here. Below we will use the term PTC {\it laser}, taking in mind that we consider it at low pump, in the LED regime.} The discussion of the results is presented in Section IV. The conclusion finalises the paper. 
\section{\label{Sec2}Model of the laser with PTC}
\subsection{\label{subec2_0}Scheme of the laser with PTC}
\begin{figure}[thb]\bc
\centering
\includegraphics[width=9.5cm]{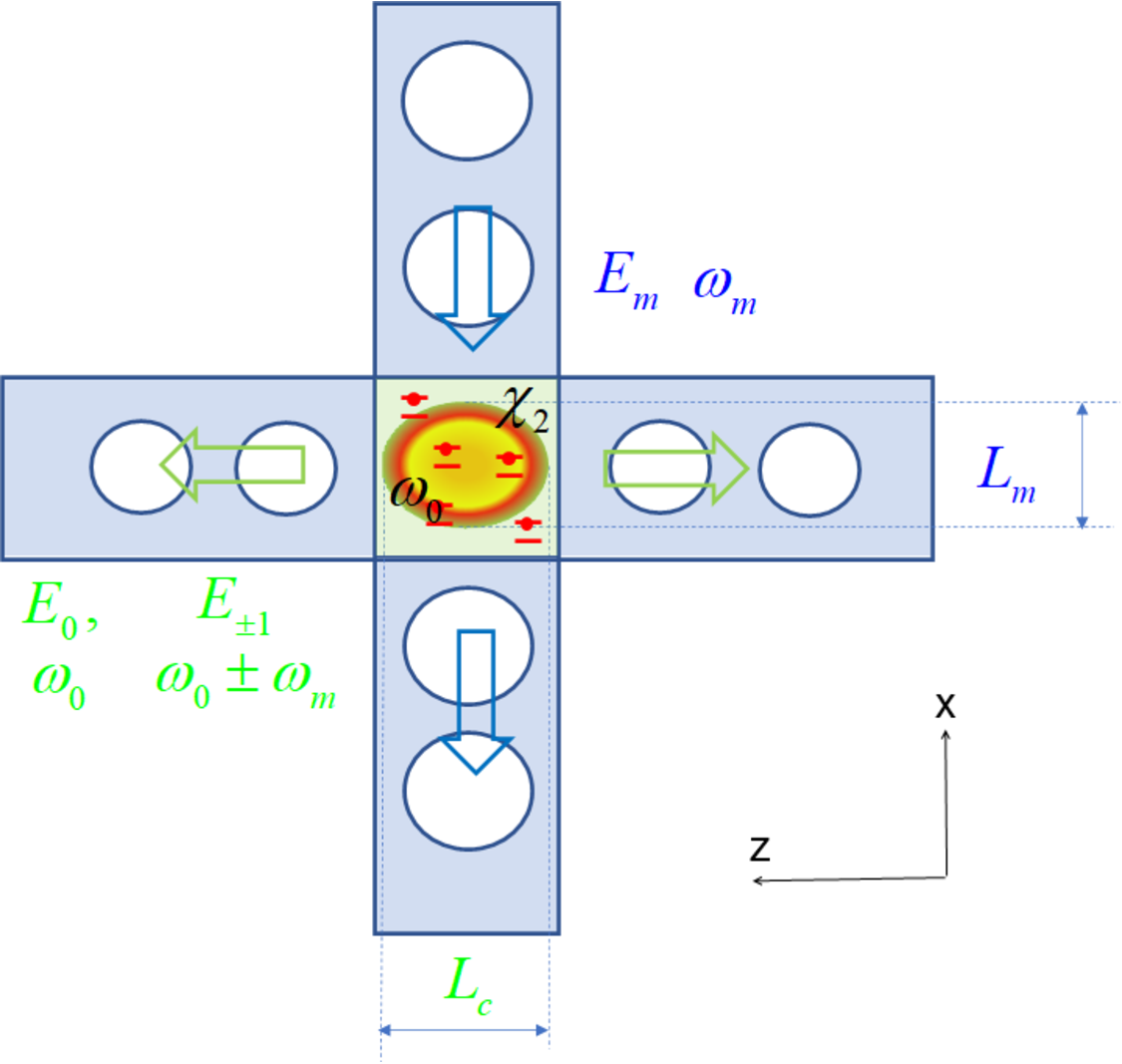}
\caption{Scheme of the micro-laser with PTC.  }
\label{Fig1}\ec
\end{figure}

We consider two intersecting one-dimensional photonic crystals \ct{Zabolotin:11}, which are, for example,  two lines of holes in dielectric slabs \ct{Okayama2024}, as shown in Figure~\ref{Fig1}. 
%
%%%%%%%%%%%%%%%%%%%%%%%%%%%%%%%%%%%%%%%%%%%%%%%
%
%
%%%%%%%%%%%%%%%%%%%%%%%%%%%%%%%%%%%%%%%%%%%%%%%
At the intersection, one of the holes in the crystals is absent, and creating a small "photonic wire" cavity  {\ct{Foresi1997,Ahmad_phd}} at the intersection. {The holes in the dielectric slabs surrounding the cavity in Figure~\ref{Fig1} form a semitransparent cavity  mirrors (Bragg mirrors) providing the electromagnetic field standing waves in the vertical and in the horizontal directions in the cavity. Figure~\ref{Fig1} presents   possible example of the cavity, other micro-cavities with the vertical and horizontal modes are possible.  }
The cavity has dimensions $L_m$ and $L_c$ in the vertical and horizontal directions, respectively, and is resonant to optical fields with frequencies $\omega_c$  and $\omega_m$ propagating in the vertical ($x$) and horizontal ($z$) crystals. In Section~\ref{Sec3} we make the following assumptions for the estimation of the defect cavity lengths: $L_c\sim \tilde{\lambda}_c/2$ with  {$\tilde{\lambda}_c = 2\pi c/(n_r\omega_c)$} and $L_m\sim \tilde{\lambda}_m/2$ with  {$\tilde{\lambda}_m = 2\pi c/(n_r\omega_m)$}, where $n_r$ is the linear refractive index inside the cavity, $c$ is the speed of light in vacuum. 

It is assumed that $N_0$  two-level active emitters, (as quantum dots) with the transition frequency $\omega_0$ are placed in the cavity and lead to lasing in the cavity. Also a  nonlinear medium with the second-order dielectric susceptibility $\chi_2$ occupies the cavity.  The field of the frequency $\omega_m$, propagating over the vertical crystal, is the external (pump) field, which modulates the nonlinear medium at the frequency  $\omega_m$, creating the photonic time crystal \ct{doi:10.1021/acsphotonics.4c00607} in the cavity. This ultimately leads to PTC effects in lasing (see below). The emitters can be positioned within a semiconductor layer located just beneath the photonic crystal \ct{Katsuno_2024}. 

Thus, we have both the resonant active medium and the PTC within the same small cavity.  
We assume that the field polarizations are perpendicular to the plane of Figure~\ref{Fig1} and dropping vector notation for the field and polarization variables. {We assume, that the cavity supports two fundamental field modes, of the minimum possible size, propagating in the horizontal and in the vertical directions. We call these modes the horisontal mode (HM) and the vertical mode (VM).}  
The two-level active emitters  generate the radiation of the  frequency, $\omega_0$.    This radiation  is resonant to {HM}.  The external field of frequency $\omega_m$ enters {the cavity in the vertical direction, it is resonant to VM} and, together with the field of frequency $\omega_0$, leads to oscillations of the nonlinear medium polarization of PTC. Under certain conditions such  
polarization oscillations generate fields of frequencies  $\omega_0\pm p\omega_m$, ($p=1,2...$), that are in resonance with the {HM} and propagate in the horizontal direction together with the field of frequency $\omega_0$. This is similar to the situation in the usual PTC without an active medium \ct{Asgari:24,Lustig:23,Boltasseva:24}. In consideration below we assume that the fields of frequencies $\omega_0\pm p\omega_m$, where $p\geq 2$, are weak and neglect them. This is an 
approximation often used in PTC theory  {\ct{Asgari:24}}. 
We then determine the conditions under which the fields of the frequencies $\omega_0$  and $\omega_0\pm p\omega_m$  coexist and are in resonance with {HM}.
For simplicity, we do not consider here  parametric resonances, when $\omega_0$ is close to $l\omega_m$, $l=1,2,...$. PTC at parametric resonances, was considered in \ct{doi:10.1021/acsphotonics.4c00607}.

Note that the {VM} in Fig.~\ref{Fig1}  allows one to resonantly increase the external field, which modulates the nonlinear medium, and correspondingly, thereby enhancing the PTC effects in {HM} lasing.  {{VM} provides the spatially independent phase for the external field (standing wave in {VM}) and the nonlinear medium polarization oscillations, which improves the coupling of the nonlinear medium polarization and the lasing field in   {HM}.}
\subsection{\label{subec2_A}The lasing mode and the laser equations}
The lasing field mode in a laser with PTC differs significantly from that in a laser without PTC. Therefore, we will provide a through analysis of the lasing mode, beginning with the Maxwell-Bloch equations. Our approach will be based on the well-known method used for single-mode lasers \ct{SLW}. First, let us derive the equations for the field in the  cavity. The electric field in the   {HM} is a standing wave
\beq
\tilde{\mathcal{E}}(\vec{r},t)=\mathcal{E}(t)R_Z(z)R_{XY}(x,y) \lb{El_F_1}
\eeq
where the spatial dependence $R_Z(z)R_{XY}(x,y)$ corresponds to the $  {HM}$  of the cross-section $S$, ${S}^{-1}\int\limits_{S}{{{R}_{XY}^{2}}(x,y)ds}=1$.  
Below we will approximate $R_Z(z)$ as $R_Z(z)\approx \sin{(2\pi z/\tilde{\lambda}_c)}$. 
The field in the   {VM} of the cross-section, $S_m$, is a standing wave 
\beq
\tilde{\mathcal{E}}_m(\vec{r},t)=E_m\cos{(\omega_m t)}R_{m,X}(x)R_{m,YZ}(y,z), \lb{VR_field}
\eeq
${S_m}^{-1}\int\limits_{S_m}{{{R}_{m,YZ}^{2}}(y,z)ds}=1$. 
  {The value of the field amplitude $E_m$ is determined by the external (pump) field.} Analogously, we will approximate $R_{mX}(x)\approx\sin{(2\pi x/\tilde{\lambda}_m)}$. We assume that spatial configurations of   {HM} and   {VM}  are known and they can be adjusted properly to achieve the correct resonant frequency and the field of the   {HM} (  {VM}).  
Polarization ${P}_2(\vec{r},t)$ of the second-order nonlinear medium in the cavity is
\beq
{P}_2(\vec{r},t)={{\chi }_{2}}{{\left[ \tilde{\mathcal{E}}(\vec{r},t)+{{{\tilde{\mathcal{E}}}}_{m}}(\vec{r},t) \right]}^{2}}, \lb{Nonl_pol_tot}
\eeq
 {where $\chi_2$ is the second-order dielectric susceptibility coefficient}. Let us consider various terms in  Eq.~\rf{Nonl_pol_tot}. The term proportional to ${\tilde{\mathcal{E}}}_{m}^{2}(\vec{r},t)$ corresponds to the second harmonic of the external electric field of frequency $\omega_m$. Some part of this second harmonic reaches the   {HM}, but the second harmonic has the frequency $2\omega_m$ and does not affect the lasing in   {HM} occurred on different frequencies, so we neglect the term proportional to ${\tilde{\mathcal{E}}}_{m}^{2}(\vec{r},t)$ in Eq.~\rf{Nonl_pol_tot}.  The external field modulating the PTC is large, ${{\tilde{\mathcal{E}}}_{m}}(\vec{r},t)>>\tilde{\mathcal{E}}(\vec{r},t)$, so the contribution of ${{\tilde{\mathcal{E}}}^{2}}(\vec{r},t)$ to ${P}_2(\vec{r},t)$ is small compared to the contributions of terms proportional to  ${{\tilde{\mathcal{E}}}{\tilde{\mathcal{E}}}_{m}}$ and ${{\tilde{\mathcal{E}}}_{m}^2}$. Therefore, we can neglect the contribution of ${{\tilde{\mathcal{E}}}^{2}}(\vec{r},t)$ to ${P}_2(\vec{r},t)$. So, we can write 
\beq
{P}_2(\vec{r},t)\approx{{\tilde{P}}_{m}}(\vec{r},t)=2{{\chi }_{2}}\tilde{\mathcal{E}}(\vec{r},t){{E}_{m}}\cos \left( {{\omega }_{m}}t \right)R_{m,X}(x){{R}_{m,YZ}}(y,z). \lb{Nonlin_P}
\eeq
The electric field $\tilde{\mathcal{E}}(\vec{r},t)$ of the   {HM} is described by the equation 
\beq
\frac{{{\partial }^{2}}\tilde{\mathcal{E}}(\vec{r},t)}{\partial {{t}^{2}}}+2\kappa \frac{\partial \tilde{\mathcal{E}}(\vec{r},t)}{\partial t}+\omega _c^{2}\tilde{\mathcal{E}}(\vec{r},t)=-4\pi \left( \frac{{{\partial }^{2}}{\tilde{P}_{m}}}{\partial {{t}^{2}}}+\frac{{{\partial }^{2}}{{{\tilde{P}}}_{e}}}{\partial {{t}^{2}}} \right). \lb{Max_no_sours}
\eeq
where $\omega_c$ is the resonance frequency of the   {HM}  (see above), $\kappa$ describes losses in   {HM} and $\tilde{P}_e$ is the polarization of the active emitters.  {Linear refractive index of the medium in   {HM} is included to $\tilde{P}_m$ and $\tilde{P}_{e}$}. Eq.~\rf{Max_no_sours}  is an oscillator-type equation that can easily be obtained from the Maxwell equations  by the same procedure as in the standard laser theory \ct{SLW}.   {Eq.~\rf{Max_no_sours} is, in fact, Mathieu-like parametric oscillator equation with the coefficients modulated at the frequency  $\omega_m$.} 
Below in Section~\ref{subec2_B} we find the frequencies  {$\omega_c$} corresponding to the resonance of the field $\tilde{\mathcal{E}}(\vec{r},t)$  to the   {HM}. 

{In accordance with Floquet theorem \ct{doi:10.7566/JPSJ.94.111007}}  we represent the time-dependent part $\mathcal{E}(t)$ of the field, $\tilde{\mathcal{E}}(\vec{r},t)$, shown by   Eq.~\rf{El_F_1}, as follows:
\beq
\mathcal{E}(t)=\sum\limits_{p=0,\pm 1,...}{{{E}_{p}}(t){{e}^{-i({{\omega }_{0}}+p{{\omega }_{m}})t}}},\lb{PTC_exp}
\eeq
where $p=0,\pm 1, \pm 2, ...$. Representation \rf{PTC_exp} is a common feature of PTC theory   \ct{Asgari:24}. We only consider terms with $p=0,\pm 1$, and disregard the others in Eq.~\rf{PTC_exp}. 

We suppose that the Fourier components 
${{E}_{0}}(\omega )$ and ${{E}_{\pm 1}}(\omega )$ of the amplitudes ${{E}_{0}}(t)$ and ${{E}_{\pm 1}}(t)$ of the fields with 
frequencies $\omega_0$ and $\omega_0\pm\omega_m$ are significantly different from zero only near the resonances at that 
frequencies. In other words,  $\omega$ is small respectively to  $\omega_0$ and $\omega_0\pm\omega_m$, therefore the field amplitudes $E_0(t)$ and $E_{\pm 1}(t)$ vary slowly, respectively, to $\exp{(-i\omega_0t)}$ and $\exp{[-i(\omega_0\pm\omega_m)t]}$. This assumption will be confirmed later when we find the spectra of the $E_0$ and $E_{\pm 1}$ fields as narrow peaks centered on the carrier frequencies, $\omega_0$ and  $\omega_0\pm\omega_m$. We substitute expression \rf{PTC_exp} into Eq.~\rf{El_F_1}, then Eq.~\rf{El_F_1} into Eq.~\rf{Max_no_sours}. We then multiply Eq.~\rf{Max_no_sours} by the spatial part,  {$ R_Z(z)R_{XY}(x,y)$}, of the   {HM}  and integrate over the volume $V$ of  {  {HM}}. By equating the terms $\sim\exp{(-i\omega_0t)}$, and $\sim\exp{[-i(\omega_0\pm\omega_m)t]}$, we arrive at a set of equations for the slowly varying amplitudes $E_0(t)$ and $E_{\pm 1}(t)$ of the  electric field in the   {HM} 
\begin{subequations}\lb{Three_ME}
\beqr
  &\left( \omega _{c}^{2}-\omega _{0}^{2}-2i{{\omega }_{0}}\kappa  \right){{E}_{0}}-2i{{\omega }_{0}}{{{\dot{E}}}_{0}}-2\pi \delta \chi \omega _{0}^{2}\left( {{E}_{1}}+{{E}_{-1}} \right)=4\pi \omega _{0}^{2}{{P}_{e}}& \lb{Three_ME_1}\\
 &\left[ \omega _{c}^{2}-{{\left( {{\omega }_{0}}+{{\omega }_{m}} \right)}^{2}}-2i\left( {{\omega }_{0}}+{{\omega }_{m}} \right)\kappa  \right]{{E}_{1}}-2i\left( {{\omega }_{0}}+{{\omega }_{m}} \right){{{\dot{E}}}_{1}}-2\pi \delta \chi {{\left( {{\omega }_{0}}+{{\omega }_{m}} \right)}^{2}}{{E}_{0}}=0 &\lb{Three_ME_2}\\ 
 &\left[ \omega _{c}^{2}-{{\left( {{\omega }_{0}}-{{\omega }_{m}} \right)}^{2}}-2i\left( {{\omega }_{0}}-{{\omega }_{m}} \right)\kappa  \right]{{E}_{-1}}-2i\left( {{\omega }_{0}}-{{\omega }_{m}} \right){{{\dot{E}}}_{-1}}-2\pi \delta \chi {{\left( {{\omega }_{0}}-{{\omega }_{m}} \right)}^{2}}{{E}_{0}}=0 & \lb{Three_ME_3} 
\eeqr\end{subequations}
Here, ${{P}_{e}}$ is the slowly varied amplitude of the  polarization of two-level emitters,  ${{\tilde{P}}_{e}}={{P}_{e}}{{e}^{-i{{\omega }_{0}}t}}$, see ${{\tilde{P}}_{e}}$ in Eq.~\rf{Max_no_sours}.  In Eqs.~\rf{Three_ME}, we neglect the slowly varying second-time derivatives of the field amplitudes and by all time derivatives of $P_{e}$, as is common in the laser theory \ct{SLW,Scully}.   {Note that the coupled equations for amplitudes  $E_0$, $E_1$  and $E_{-1}$  resemble the coupled mode equations in consideration, at least, of some schemes of SFG/DFG \ct{boyd2020nonlinear}.} The normalized second-order dielectric susceptibility coefficient is
\beq
\delta \chi =4C{{\chi }_{2}}{{E}_{m}}, \lb{delta_chi}
\eeq
where
\beq
C=\frac{1}{V}\int\limits_{V}R_Z(z)R_{m,X}(x)R_{XY}(x,y){{R}_{m,YZ}}(y,z)dxdydz \lb{coef_C}
\eeq
represents the overlap of the   {HM} and   {VM}  in space. 

It is convenient to describe the polarization of two-level active medium in terms of the two-level system operators, similar as it is in the semiclassical laser theory \ct{SLW}. The  polarization of the $N_0$ two-level emitters is  ${{P}_{e}}=d\sum\limits_{i=1}^{{{N}_{0}}}{{{{\hat{\sigma }}}_{i}}\delta (\vec{r}-{{{\vec{r}}}_{i}})}$, where $d$ is the matrix element of the dipole  transition of the two-level emitter,  $\delta (\vec{r}-\vec{r}_i)$ is the density corresponding to a single emitter and $\hat{\sigma}_i$ is the lowering operator, describing the transition from the upper to the lower state of the i-th emitter. We multiply the left-hand side of Eq.~\rf{Max_no_sours} by  {$(1/V)R_Z(z)R_{XY}(x,y)$} and integrate over $V$. After integrating over $V$, we have in  Eqs.~\rf{Three_ME}
\beq
{{P}_{e}}=\frac{2d}{V}\sum\limits_{i=1}^{{{N}_{0}}}{{{f}_{i}}{{{\hat{\sigma }}}_{i}}}, \lb{P_NL_1}
\eeq
where  {${{f}_{i}}=R_Z(z_i)R_{XY}(x_i,y_i)$} and $x_i$, $y_i$, $z_i$ are the coordinates of the i-th emitter. 

Let us use Eq.~\rf{P_NL_1} to write  Eq.~\rf{Three_ME_1} in the form
\beq
\left[ i\left( \omega_{c}^{2}-\omega _{0}^{2} \right)/2{{\omega }_{0}}+\kappa  \right]{{E}_{0}}+{{\dot{E}}_{0}}-i\pi \delta \chi {{\omega }_{0}}\left( {{E}_{1}}+{{E}_{-1}} \right)=\frac{4\pi i{\omega }_{0}d}{V}\sum\limits_{i=1}^{{{N}_{0}}}{{{f}_{i}}{{{\hat{\sigma }}}_{i}}} \lb{First_eq}
\eeq
We now express the amplitudes $E_0$ and $E_{\pm 1}$ of the fields in the   {HM} in terms of the Bose-operators $\hat{a}_0$, $\hat{a}_{\pm 1}$ 
\beq
{{E}_{0}}={{E}_V}{{\hat{a}}_{0}}, \hspace{0.5cm} {{E}_{\pm 1}}={{E}_V}{{\left( 1\pm {{\omega }_{m}}/{{\omega }_{0}} \right)}^{1/2}}{{\hat{a}}_{\pm 1}}, \lb{field_BO}
\eeq
where ${{E}_V}={{\left( 4\pi \hbar {{\omega }_{0}}/V \right)}^{1/2}}$. The field amplitude Bose operators given by Eq.~\rf{field_BO} will be helpful in the calculations of the field spectra in sections~\ref{Sec_PNS} and \ref{subec2_C}. We substitute expression \rf{field_BO} into equation \rf{First_eq} to derive an equation that connects the field and the polarization amplitude operators $\hat{a}_0$ and $\hat{v}=i\sum\limits_{i=1}^{{{N}_{0}}}{{{f}_{i}}{{{\hat{\sigma }}}_{i}}}$
\beq
\left[ i\left( \omega _{c}^{2}-\omega _{0}^{2} \right)/2{{\omega }_{0}}+\kappa  \right]{{\hat{a}}_{0}}+{{\dot{\hat{a}}}_{0}}-i\pi \delta \chi {{\omega }_{0}}\left[ {{\left( 1+{{\omega }_{m}}/{{\omega }_{0}} \right)}^{1/2}}{{{\hat{a}}}_{1}}+{{\left( 1-{{\omega }_{m}}/{{\omega }_{0}} \right)}^{1/2}}{{{\hat{a}}}_{-1}} \right]=\Omega \hat{v},\lb{quant_Eq1}
\eeq
where $\Omega ={{\left( 4\pi \hbar {{\omega }_{0}}/V \right)}^{1/2}}\left( d/\hbar  \right)$ is the vacuum Rabi frequency.  In a similar way, we can derive the equations for the field amplitude operators $\hat{a}_{\pm 1}$ by substituting  Eqs.~\rf{field_BO} into Eqs.~\rf{Three_ME_2} and \rf{Three_ME_3}. We introduce normalized detunings ${{\delta }_{0}}=\left( \omega _{c}^{2}-\omega _{0}^{2} \right)/2{{\omega }_{0}}$, ${{\delta }_{\pm 1}}=\left[ \omega _{c}^{2}-{{\left( {{\omega }_{0}}\pm {{\omega }_{m}} \right)}^{2}} \right]/2\left( {{\omega }_{0}}\pm {{\omega }_{m}} \right)$ and normalized nonlinear coupling rates ${{h}_{\pm 1}}=\pi \delta \chi {{\omega }_{0}}{{\left( 1\pm {{\omega }_{m}}/{{\omega }_{0}} \right)}^{1/2}}$, rewrite Eqs.~\rf{Three_ME} in the new notations, and add the Langevin forces to Eqs.~\rf{Three_ME}, as described in references  \ct{Andre:19,Protsenko_2021,PhysRevA.105.053713}. This leads to a set of equations for operators $\hat{a}_0$ and $\hat{a}_{\pm 1}$
\begin{subequations}\lb{q_MB}
\beqr
  {{{\dot{\hat{a}}}}_{0}}&=&-\left( i{{\delta }_{0}}+\kappa  \right){{{\hat{a}}}_{0}}+i\left( {{h}_{1}}{{{\hat{a}}}_{1}}+{{h}_{-1}}{{{\hat{a}}}_{-1}} \right)+\Omega \hat{v}+\sqrt{2\kappa }{{{\hat{a}}}_{0in}} \lb{q_MB_1} \\
 {{{\dot{\hat{a}}}}_{1}}&=&-\left( i{{\delta }_{1}}+\kappa  \right){{{\hat{a}}}_{1}}+i{{h}_{1}}{{{\hat{a}}}_{0}}+\sqrt{2\kappa }{{{\hat{a}}}_{1in}} \lb{q_MB_2} \\ 
  {{{\dot{\hat{a}}}}_{-1}}&=&-\left( i{{\delta }_{-1}}+\kappa  \right){{{\hat{a}}}_{-1}}+i{{h}_{-1}}{{{\hat{a}}}_{0}}+\sqrt{2\kappa }{{{\hat{a}}}_{-1in}} 
\lb{q_MB_3}
\eeqr\end{subequations}
Note that Heisenberg equations~\rf{q_MB} can be obtained from Hamiltonian
\beq
H=\hbar {{\delta }_{0}}\hat{a}_{0}^{+}{{\hat{a}}_{0}}+\hbar {{\delta }_{1}}\hat{a}_{1}^{+}{{\hat{a}}_{1}}+\hbar {{\delta }_{-1}}\hat{a}_{-1}^{+}{{\hat{a}}_{-1}}-\hbar {{h}_{1}}\left( \hat{a}_{1}^{+}{{{\hat{a}}}_{0}}+\hat{a}_{0}^{+}{{{\hat{a}}}_{1}} \right)-\hbar {{h}_{-1}}\left( \hat{a}_{-1}^{+}{{{\hat{a}}}_{0}}+\hat{a}_{0}^{+}{{{\hat{a}}}_{-1}} \right)+\hat{V}_{int}+\hat{\Gamma }\lb{HAM}
\eeq
where $\hat{V}_{int} = i\hbar\Omega(\hat{a}_0^+\hat{v} - \hat{v}^+\hat{a}_0)$ describes the interaction of the active medium with  the   {HM}  at the carrier frequency $\omega_0$, and the term $\hat{\Gamma}$ describes the field losses in the cavity. The dissipative terms $\sim \kappa$ and the Langevin forces $\sim \sqrt{2\kappa}$ with the vacuum field Bose operators   {$\hat{a}_{0,\pm1in}$} describe the losses of the cavity field in Eqs.~\rf{q_MB}.

The active medium of two-level emitters interacts with the field  of frequency $\omega_0$ in the same way as in the  laser without PTC. The equations for the operators $\hat{v}$ of the active medium polarization and upper state population $\hat{N}_e$  are the same as those for a laser without PTC  \ct{Scully,Andre:19,Protsenko_2021,PhysRevA.105.053713}
\begin{subequations}\lb{ac_m_eq}
\beqr
  \dot{\hat{v}} &=&-\left( {{\gamma }_{\bot }}/2 \right)\hat{v}+\Omega f{{{\hat{a}}}_{0}}\left( {{{\hat{N}}}_{e}}-{{{\hat{N}}}_{g}} \right)+{{{\hat{F}}}_{v}}, \lb{ac_m_eq_1}\\ 
{{{\dot{\hat{N}}}}_{e}}&=&-\Omega \left( \hat{a}_{0}^{+}\hat{v}+{{{\hat{v}}}^{+}}{{{\hat{a}}}_{0}} \right)+{{\gamma }_{\parallel }}\left( P{{{\hat{N}}}_{g}}-{{{\hat{N}}}_{e}} \right)+{{{\hat{F}}}_{{{N}_{e}}}}\lb{ac_m_eq_2}
\eeqr\end{subequations}
where $\hat{N}_g$ is the operator of the population of the low emitter states, $\hat{N}_e+\hat{N}_g=N_0$, and $\gamma_{\perp}$, $\gamma_{\parallel}$ and $P\gamma_{\parallel}$ are, respectively, the linewidth of the  of two-level emitter transition (polarization relaxation rate) and the upper emitter level dumping and pump rates, respectively. ${{{\hat{F}}}_{v}}$ and ${{{\hat{F}}}_{{{N}_{e}}}}$ are Langevin forces and  {$f=N_0^{-1} \sum_{i=1}^{N_0}f_i^2\approx 1/2$}. The  correlation of the Langevin force 
\beq
\left<\hat{F}_{v^+}(t)\hat{F}_{v}(t')\right> = f\gamma_{\perp}N_e\delta(t-t') \lb{corr_LF_v} 
\eeq
will be needed in the calculations below. The mean populations $N_{e,g}$ of the emitter states can be found from  the  {relation}
\beq
 {2\kappa\left( {{n}_{0}}+{{n}_{1}}+{{n}_{-1}} \right)={{\gamma }_{\parallel }}\left( P{{N}_{g}}-{{N}_{e}} \right)}, \lb{eql}
\eeq
 {followed from Eqs.~\rf{q_MB}, \rf{ac_m_eq}};  ${{n}_{i}}=\left\langle \hat{a}_{i}^{+}{{{\hat{a}}}_{i}} \right\rangle $, $i=0,\pm 1$ is the mean photon number for each of the three carrier frequencies.  {Deriving Eq.~\rf{eql} we write equations for $\hat{a}_{i}^{+}{{{\hat{a}}}_{i}}$ from Eqs.~\rf{q_MB}, sum them and see that  all terms $\sim i$ are canceled in the sum. Then we take the mean value of the sum, express $\Omega(\left<\hat{a}_0^+\hat{v}\right> + c.c)={{\gamma }_{\parallel }}\left( P{{{{N}}}_{g}}-{{{{N}}}_{e}} \right)$ using the stationary  Eq.~\rf{ac_m_eq_2} and come to Eq.~\rf{eql}.

Eq.~\rf{eql} means that the lasing photon decay rate is equal to the active medium excitation rate. Eq.~\rf{eql} is a kind of a conservation relation for observables in Folquet quantum systems,   here -- for the PTC laser with constant modulation field. See some examples of quasi-conserved observables in Floquet quantum systems in \ct{ PhysRevB.103.054305}. In a PTC laser, the excitation of the active medium by the quant $\hbar\omega_0$ produces field quanta of frequency $\hbar\omega_0$ or $\hbar(\omega_0\pm\omega_m)$ in the lasing mode. The energy mismatch caused by the difference between the excitation of the active medium, $\hbar\omega_0$, and the quanta of the field, $\hbar(\omega_0\pm\omega_m)$, must be compensated, for example, by the energy exchange through the nonlinear medium between the modulation and lasing fields. In the present approximation, where a constant modulation field is assumed, we cannot consider such an energy exchange and the modulation field variation. We leave the analysis of the modulation field dynamics for consideration in a future paper.}  

Following \ct{Andre:19,Protsenko_2021,PhysRevA.105.053713}, we find the stationary solutions of Eqs.~\rf{q_MB}, \rf{ac_m_eq} by neglecting the active medium population fluctuations.  
\subsection{\label{subec2_B}Conditions for resonant lasing in the cavity with PTC}
Below we find conditions for resonance generation in the   {HM of the cavity} with PTC. This is done in a similar way to a laser without PTC. For convenience, we use Eqs.~\rf{Three_ME} for the field amplitudes.

In the absence of a nonlinear medium, the   {HM interacting} with an active medium behaves as   {in} a usual laser, when the cavity mode field amplitude of which $E_0$ satisfies equation 
\beq
\left( \omega_{c}^{2}-\omega _{0}^{2}-2i{{\omega }_{0}}\kappa  \right){{E}_{0}}-2i{{\omega }_{0}}{{\dot{E}}_{0}}=4\pi \omega _{0}^{2}{{P}_{e}}. \lb{MB_eq_usual}
\eeq
The field losses term $\sim \kappa$, the active medium polarization term $\sim P_{e}$, and the slowly varied in time term $\sim \dot{E}_0$ are neglected  when calculating $\omega_c$  \ct{SLW,Scully}. We write a simple equation for $\omega_c^{2}$, 
\beq
\left( \omega _{c}^{2}-\omega _{0}^{2}  \right){{E}_{0}}=0, \lb{Tr_vawen_eq}
\eeq
when we neglect $\kappa$, $P_{e}$ and $\dot{E}_0$ in Eq.~\rf{MB_eq_usual}. Equation \rf{Tr_vawen_eq} has a non-trivial solution ${{E}_{0}}\neq 0$ if $\omega _{c}^{2}=\omega _{0}^{2}$.  
Thus, in a conventional laser cavity without a PTC, lasing occurs at the resonance frequency  {$\omega_c$} of the cavity.  

We generalize this approach to a laser with a PTC for which three equations \rf{Three_ME} describe the cavity  field. By neglecting the loss term proportional to $\kappa$, and the term involving the active medium polarization $P_{e}$, we obtain a set of algebraic equations for the field amplitudes  $E_0$ and $E_{\pm 1}$:
\begin{subequations}\lb{eigen_mode}
\beqr
 \left( \omega _{c}^{2}-\omega _{0}^{2} \right){{E}_{0}}-2\pi \delta \chi \omega _{0}^{2}\left( {{E}_{1}}+{{E}_{-1}} \right)&=&0 \lb{eigen_mode_1} \\ 
 \left[ \omega _{c}^{2}-{{\left( {{\omega }_{0}}+{{\omega }_{m}} \right)}^{2}} \right]{{E}_{1}}-2\pi \delta \chi {{\left( {{\omega }_{0}}+{{\omega }_{m}} \right)}^{2}}{{E}_{0}}&=&0\lb{eigen_mode_2} \\ 
 \left[ \omega _{c}^{2}-{{\left( {{\omega }_{0}}-{{\omega }_{m}} \right)}^{2}} \right]{{E}_{-1}}-2\pi \delta \chi {{\left( {{\omega }_{0}}-{{\omega }_{m}} \right)}^{2}}{{E}_{0}}&=&0 \lb{eigen_mode_3} 
\eeqr\end{subequations}
In fact, Eqs.~\rf{eigen_mode} generalize Eq.~\rf{Tr_vawen_eq} to the case of a laser cavity with PTC. Eqs.~\rf{eigen_mode} have a nontrivial solution for the amplitudes  if the determinant of Eqs.~\rf{eigen_mode} is zero: 
\beqr
  & \left( \omega _{c}^{2}-\omega _{0}^{2} \right)\left[ \omega _{c}^{2}-{{\left( {{\omega }_{0}}+{{\omega }_{m}} \right)}^{2}} \right]\left[ \omega _{c}^{2}-{{\left( {{\omega }_{0}}-{{\omega }_{m}} \right)}^{2}} \right]- &\lb{char_eq}\\ 
 & {{\left( 2\pi \delta \chi {{\omega }_{0}} \right)}^{2}}\left\{ {{\left( {{\omega }_{0}}-{{\omega }_{m}} \right)}^{2}}\left[ \omega _{c}^{2}-{{\left( {{\omega }_{0}}+{{\omega }_{m}} \right)}^{2}} \right]+{{\left( {{\omega }_{0}}+{{\omega }_{m}} \right)}^{2}}\left[ \omega _{c}^{2}-{{\left( {{\omega }_{0}}-{{\omega }_{m}} \right)}^{2}} \right] \right\}=0. & \nonumber
\eeqr
Eq.~\rf{char_eq} establishes the relationship between the frequencies $\omega_0$, $\omega_c$  and $\omega_m$  and determines the lasing regimes in the cavity with PTC. In particular, one might consider that Eq.~\rf{char_eq} determines the resonance frequency $\omega_c$  of the cavity at given frequencies $\omega_0$, and $\omega_m$  (see Fig.~\ref{Fig2} below, for example). 
Obviously, at $\delta\chi\rightarrow 0$ we have three regimes (modes) for lasing in the cavity:
$\omega_c = \omega_0$, 
$\omega_c =\omega_0+\omega_m$ and
$\omega_c =\omega_0-\omega_m$. For convenience, these frequencies can be denoted as $\omega_{c,p} =\omega_0\pm p\omega_m$  where $p=0,\pm 1$ (see also Fig.~\ref{Fig2}). It is worth noting here that we have  {three resonant frequencies of the cavity (three modes)} because we only considered two sidebands ($p=0,\pm 1$) in the   {HM} field -- see Eq.~\rf{PTC_exp}. Without this restriction, there is an infinite number of   {the cavity modes propagating in the horizontal direction} with the frequencies 
$\omega_{c,p} =\omega_0+p\omega_m$,
($p=0,\pm 1,\pm 2...$) when  $\delta\chi\rightarrow 0$. 

Let us introduce the normalized quantities 
$X={{\left( {{\omega }_{c}}/{{\omega }_{0}} \right)}^{2}}$, $X_m={{\left( {{\omega }_m}/{{\omega }_{0}} \right)}^{2}}$, and the nonlinear coupling coefficient $h=\pi \delta \chi $, and write the following instead of Eq.~\rf{char_eq}: 
\beq
\left( X-1 \right)\left[ {{X}^{2}}-2X\left( 1+{{X}_{m}} \right)+{{\left( 1-{{X}_{m}} \right)}^{2}} \right]-8{{h}^{2}}\left[ X\left( 1+{{X}_{m}} \right)-{{(1-{{X}_{m}})}^{2}} \right]=0.\lb{char_eq_norm}
\eeq
Equation~\rf{char_eq_norm} is cubic with respect to $X$ and has three real roots at positive  $X_m$ and $h$.  The   {HM} which can exist in the   {cavity} with PTC correspond to positive roots of Eq.\rf{char_eq_norm}. 

Let us consider what happens when one of the roots of equation~\rf{char_eq_norm} becomes zero, changing its sign from positive to negative. Suppose that we  find that one of the roots of Eq.~\rf{char_eq_norm} becomes zero at a given values of $X_m$ and $h$. If one of the roots of Eq.~\rf{char_eq_norm} is zero, the term without $X$ in Eq.~\rf{char_eq_norm} is also zero. This means that
\beq
(8{{h}^{2}}-1){{(1-{{X}_{m}})}^{2}}=0.\lb{free_term_zero}
\eeq
From Eq.~\rf{free_term_zero}, one can see that one of the roots of Eq.~\rf{char_eq_norm} changes the sign and becomes zero when $8h^2=1$ or $X_m=1$. For certainty, we  assume that the PTC modulation frequency, $\omega_m$, is smaller than the lasing transition frequency, $\omega_0$, so that $X_m\equiv(\omega_m/\omega_0)^2<1$.   For $0<8h^2<1$, we can see that all three roots of Eq.~\rf{char_eq_norm} are positive. Therefore,  one of   {HM} corresponds to $X<0$ when $8h^2>1$.   

Note that above link between the frequencies $\omega_0$, $\omega_c$  and $\omega_m$  can be interpreted also as following. Nonlinear modulation of PTC inside the cavity with the “bare” resonance frequency $\omega_c$  leads to additional “dressing” modes (side modes) in the cavity with the frequencies $\omega_c\pm\omega_m$  (in the limit $\delta\chi\rightarrow 0$). Correspondingly, the lasing in the “dressed” cavity can occur at the transition frequencies $\omega_0\approx\omega_c$  and $\omega_0\approx\omega_c\pm\omega_m$.
\subsection{\label{subec2_D}Slowly varying field amplitudes and relations between field spectra}\label{Sec_PNS}
We will find slowly varying field amplitudes, $E_{0\pm 1}$, at resonance, i.e. when $\omega_{c}^2$ is a solution of Eq.~\rf{char_eq}. To achieve this, we write from Eqs.~\rf{Three_ME}    equations for Fourier components $E_{0,\pm 1\omega}$ and  $P_{e\omega}$ of the field  $E_{0\pm 1}(t)$ and the active medium polarization  $P_{e}(t)$ amplitudes 
\begin{subequations}\lb{FC_slo}
\beqr
  \left[ i\left( \omega _{c}^{2}-\omega _{0}^{2} \right)/2{{\omega }_{0}}+\xi  \right]{{E}_{0\omega }}-ih{{\omega }_{0}}\left( {{E}_{1\omega }}+{{E}_{-1\omega }} \right)&=&2\pi i{{\omega }_{0}}{{P}_{e\omega }} \lb{FC_slo_1}\\ 
 \left\{ i\left[ \omega _{c}^{2}-{{\left( {{\omega }_{0}}+{{\omega }_{m}} \right)}^{2}} \right]/2\left( {{\omega }_{0}}+{{\omega }_{m}} \right)+\xi  \right\}{{E}_{1\omega }}-ih\left( {{\omega }_{0}}+{{\omega }_{m}} \right){{E}_{0\omega }}&=&0 \lb{FC_slo_2} \\ 
 \left\{ i\left[ \omega _{c}^{2}-{{\left( {{\omega }_{0}}-{{\omega }_{m}} \right)}^{2}} \right]/2\left( {{\omega }_{0}}-{{\omega }_{m}} \right)+\xi  \right\}{{E}_{-1\omega }}-ih\left( {{\omega }_{0}}-{{\omega }_{m}} \right){{E}_{0\omega }}&=&0 \lb{FC_slo_3} 
\eeqr\end{subequations}
In Eqs.~\rf{FC_slo}, $\xi =\kappa -i\omega$ and we neglect the slowly varied second time derivatives of the field and by all time derivatives of the polarization, as it is common in the laser theory \ct{SLW,Scully}. We assume narrow spectra $E_{0,\pm 1\omega}$ and  $P_{e\omega}$, so that $\omega \sim \kappa <<{{\omega }_m},{{\omega }_{0}}$. Therefore  $\xi$  is small respectively to $\omega_0$, $\omega_m$. We will find the solution of Eqs.~\rf{FC_slo} to the first non-vanishing order of a small $\xi$.

The main contribution to $E_{\pm1\omega}$ comes  from a zero-order on $\xi$ terms
\begin{subequations}\lb{solution_Epm1}\beqr
  {{E}_{1\omega }}&=&\frac{ih\left( {{\omega }_{0}}+{{\omega }_{m}} \right){{E}_{0\omega }}}{i\left[ \omega _{c}^{2}-{{\left( {{\omega }_{0}}+{{\omega }_{m}} \right)}^{2}} \right]/2\left( {{\omega }_{0}}+{{\omega }_{m}} \right)+\xi }\approx \frac{2h{{\left( {{\omega }_{0}}+{{\omega }_{m}} \right)}^{2}}}{\omega _{c}^{2}-{{\left( {{\omega }_{0}}+{{\omega }_{m}} \right)}^{2}}}{{E}_{0\omega }} \lb{solution_Epm1_1}\\ 
 {{E}_{-1\omega }}&=&\frac{ih\left( {{\omega }_{0}}-{{\omega }_{m}} \right){{E}_{0\omega }}}{i\left[ \omega _{c}^{2}-{{\left( {{\omega }_{0}}-{{\omega }_{m}} \right)}^{2}} \right]/2\left( {{\omega }_{0}}-{{\omega }_{m}} \right)+\xi }\approx \frac{2h{{\left( {{\omega }_{0}}-{{\omega }_{m}} \right)}^{2}}}{\omega _{c}^{2}-{{\left( {{\omega }_{0}}-{{\omega }_{m}} \right)}^{2}}}{{E}_{0\omega }} \lb{solution_Epm1_2}
\eeqr\end{subequations}
We substitute $E_{\pm1\omega}$ from Eqs.~\rf{solution_Epm1} to Eq.~\rf{FC_slo_1} and find:
\[
  \left[ i\left( \omega _{c}^{2}-\omega _{0}^{2} \right)/2{{\omega }_{0}}+\xi  \right]{{E}_{0\omega }}-\frac{2i{{h}^{2}}{{\omega }_{0}}{{\left( {{\omega }_{0}}+{{\omega }_{m}} \right)}^{2}}}{\omega _{c}^{2}-{{\left( {{\omega }_{0}}+{{\omega }_{m}} \right)}^{2}}}\left( 1+\frac{2i\left( {{\omega }_{0}}+{{\omega }_{m}} \right)\xi }{\omega _{c}^{2}-{{\left( {{\omega }_{0}}+{{\omega }_{m}} \right)}^{2}}} \right){{E}_{0\omega }} \]
\beq -\frac{2i{{h}^{2}}{{\omega }_{0}}{{\left( {{\omega }_{0}}-{{\omega }_{m}} \right)}^{2}}}{\omega _{c}^{2}-{{\left( {{\omega }_{0}}-{{\omega }_{m}} \right)}^{2}}}\left( 1+\frac{2i\left( {{\omega }_{0}}-{{\omega }_{m}} \right)\xi }{\omega _{c}^{2}-{{\left( {{\omega }_{0}}-{{\omega }_{m}} \right)}^{2}}} \right){{E}_{0\omega }}=2\pi i{{\omega }_{0}}{{P}_{e\omega }}\lb{Main_field} 
\eeq
If $\omega_c^2$ is the root of the equation \rf{char_eq} or \rf{char_eq_norm}, then the zero-order term in $\xi$ in Eq.~\rf{Main_field} is zero. The remaining, first-order  term in $\xi$ in Eq.~\rf{Main_field}     determines the  Fourier-component $E_{0\omega}$
\beq
\left( \kappa -i\omega  \right)R{{E}_{0\omega }}=2\pi i{{\omega }_{0}}{{P}_{e\omega }},\lb{zero_order_FC}
\eeq
where  {we replace $\xi$ by $\kappa-i\omega$ and introduce}
\beq
R=1+4{{h}^{2}}\left\{ \frac{{{\omega }_{0}}{{\left( {{\omega }_{0}}+{{\omega }_{m}} \right)}^{3}}}{{{\left[ \omega _{c}^{2}-{{\left( {{\omega }_{0}}+{{\omega }_{m}} \right)}^{2}} \right]}^{2}}}+\frac{{{\omega }_{0}}{{\left( {{\omega }_{0}}-{{\omega }_{m}} \right)}^{3}}}{{{\left[ \omega _{c}^{2}-{{\left( {{\omega }_{0}}-{{\omega }_{m}} \right)}^{2}} \right]}^{2}}} \right\}.\lb{R_coeff}
\eeq
We come to Fourier-component operators ${{\hat{a}}_{0\omega }}$ and ${{\hat{v}}_{\omega }}$ of  the field $E_{0}$ and the active medium polarization $P_{e}$ and obtain from Eqs.~\rf{zero_order_FC} and
\rf{ac_m_eq_1} equations for ${{\hat{a}}_{0\omega }}$ and ${{\hat{v}}_{\omega }}$
\begin{subequations}\lb{BE_fc}\beqr
0&=&\left( i\omega -\kappa  \right){{{\hat{a}}}_{0\omega }}+\left( \Omega /R \right){{{\hat{v}}}_{\omega }} \lb{BE_fc_1}\\ 
 0&=&\left( i\omega -{{\gamma }_{\bot }}/2 \right){{{\hat{v}}}_{\omega }}+\Omega f{{{\hat{a}}}_{0\omega }}N+{{{\hat{F}}}_{v\omega }}. \lb{BE_fc_2} 
\eeqr\end{subequations}
Here, $N=N_e-N_g$ is the mean population inversion of the active medium. In Eqs.~\rf{BE_fc}, we neglect by population fluctuations and drop the vacuum field Langevin force in Eq.~\rf{BE_fc_1}, as this does not contribute  {to the  field spectra, which we calculate}. The correlation for the polarization Langevin force Fourier component ${{{\hat{F}}}_{v\omega}}$ follows from Eq.~\rf{corr_LF_v}
\beq
\left<\hat{F}_{v^+}(\omega)\hat{F}_{v}(\omega')\right> = f\gamma_{\perp}N_e\delta(\omega-\omega'). \lb{corr_LF_v_fc} 
\eeq
Using relations \rf{field_BO} we can derive the field Bose operators from the field amplitudes   in Eqs.~\rf{solution_Epm1}. Then we apply the relation
\beq
\left<(\hat{a}^+_{0,\pm1})_{ \omega}\hat{a}_{0,\pm 1 \omega'}\right>=n_{0,\pm1\omega}\delta(\omega-\omega') \lb{field_sp_gn}
\eeq
to obtain the photon number spectra 
$n_{\pm1\omega}$ of fields with carrier frequencies $\omega_0\pm\omega_{\pm1}$. Such spectra, written in terms of the photon number spectrum $n_{0\omega}$ of the field with the carrier frequency $\omega_0$, are
\beq
{{n}_{\pm 1\omega }}={{\left[ \frac{2h{{\left( {{\omega }_{0}}\pm {{\omega }_{m}} \right)}^{2}}}{\omega _{c}^{2}-{{\left( {{\omega }_{0}}\pm {{\omega }_{m}} \right)}^{2}}} \right]}^{2}}\frac{{{n}_{0\omega }}}{\left( 1\pm {{\omega }_{m}}/{{\omega }_{0}} \right)}.\lb{n_pm1}
\eeq
The mean photon number for each carrier frequency is $n_{0,\pm1} = (2\pi)^{-1}\int_{-\infty}^{\infty}n_{0,\pm1\omega}d\omega$. The mean populations $N_{e,g}$ of the active medium  emitter states are determined from the  {relation} \rf{eql}, which can be expressed in terms of $n_0$ as follows:
\beq
 {2\kappa {{R}}{{n}_{0}}={{\gamma }_{\parallel }}\left( P{{N}_{g}}-{{N}_{e}} \right)},\lb{ecl_R}
\eeq
To complete the analysis of the laser with PTC, we need to find the field spectrum $n_{0\omega}$ and  the mean populations $N_{e,g}$. This will be done in the next section.
\subsection{\label{subec2_C}Field spectra and mean photon numbers}
Let us derive the explicit expressions for the field spectra  in   {HM}. From Eqs.~\rf{BE_fc} we find 
\beq
{{\hat{a}}_{0\omega }}=\frac{\left( \Omega /R \right){{{\hat{F}}}_{v\omega }}}{\left( i\omega -\kappa  \right)\left( i\omega -{{\gamma }_{\bot }}/2 \right)-\left( {{\Omega }^{2}}f/R \right)N}, \lb{a0_expl}
\eeq
where $R$ is given by Eq.~\rf{R_coeff}. Using Eq.~\rf{a0_expl} and the power spectrum \rf{corr_LF_v_fc} for $\hat{F}_v$, and applying  relation \rf{field_sp_gn}, we calculate the photon number spectrum of the field with the carrier frequency $\omega_0$
\beq
n_{0\omega }=\frac{1}{R}\frac{(\Omega^2f/R)N_e}{\left[\omega^2-\kappa\gamma_{\perp}/2+(\Omega^2f/R)N\right]^2+\omega^2(\kappa+\gamma_{\perp}/2)^2},\lb{n0_sp_expl}
\eeq
{From Eqs.~\rf{n0_sp_expl}  we can see that  the field-active medium coupling in PTC laser is reduced $\sqrt{R}$ times respectively to the laser without PTC: the factor $\Omega^2f$ in the photon number spectrum for the usual laser (see \ct{Andre:19,Protsenko_2021,PhysRevA.105.053713}) is replaced by the factor $\Omega^2f/R$ in Eq.\rf{n0_sp_expl} for the laser with PTC with $R>1$ (see Eq.~\rf{R_coeff}).} Consequently,  the  photon number spectrum $n_{0\omega}$ is reduced: $n_{0\omega}\sim 1/R$.

Such effects of PTC on lasing are understandable. The PTC laser field  includes three carrier frequencies, $\omega_0$, $\omega_0+\omega_m$ and $\omega_0-\omega_m$. The active medium transitions are only resonant with the frequency $\omega_0$. Additional loss channels for active medium energy are therefore created by the sideband fields of the $\omega_0\pm\omega_m$  frequencies, which interact with the field of frequency $\omega_0$ through the nonlinear medium in PTC. Such losses are not present in the laser without PTC.  These additional energy losses   {reduce the effective field-medium coupling} and the number of photons at the carrier frequency, $\omega_0$.

Integrating the spectrum \rf{n0_sp_expl} over frequencies, we find the mean photon number for the carrier frequency, $\omega_0$   {
\beq
{{n}_{0}}=\frac{1}{R}\frac{{{N}_{e}}}{\left( 1+2\kappa /{{\gamma }_{\bot }} \right)\left( R{N}_{th}-N \right)},\lb{mean_n0}
\eeq}
where $N=2N_e-N_0$ is the mean population inversion    {and $N_{th} = \kappa\gamma_{\perp}/(2\Omega^2f)$ is the threshold population inversion found in the semiclassical laser theory \ct{SLW}. We see that the effective threshold population inversion in PTC laser is increased $R$ times respectively to the usual laser without PTC; $N<RN_{th}$, $R>1$. We note that $N\rightarrow RN_{th}$ at large pump $P\gg 1$ similar to the usual laser, where the spontaneous emission to the cavity mode is taken into account   \ct{Andre:19,Protsenko_2021,PhysRevA.105.053713}.}

To find $N_e$, $n_0$ and $n_{\pm 1}$, we substitute $n_0(N_e)$ from Eq.~\rf{mean_n0} into Eq.~\rf{ecl_R}, obtaining a  quadratic equation for the normalized upper state population    {${{n}_{e}}={{N}_{e}}/{R{N}_{th}}$}:
\beq
 {{{\beta }}{{n}_{e}}=\left[ P{{{\tilde{n}}}_{0}}-\left( P+1 \right){{n}_{e}} \right]\left[ 1-\left( 2{{n}_{e}}-{{{\tilde{n}}}_{0}} \right) \right],}\lb{N_e_norm_eq}
\eeq
 {where   {${{\tilde{n}}_{0}}={{N}_{0}}/{R{N}_{th}}$}. The parameter  ${\beta }={2\kappa /{\left( 1+2\kappa /{{\gamma }_{\bot }} \right){{\gamma }_{\parallel }}}{{N}_{th}}}$ is a beta-factor \ct{PhysRevA.50.4318}.}   

Substituting ${n}_0(N_e)$ from Eq.~\rf{mean_n0} into Eq.~\rf{N_e_norm_eq} and solving Eq.~\rf{N_e_norm_eq} for $N_e$, we find the stationary mean population of the upper states of the emitters
\beq
{{N}_{e}}=\frac{{{N}_{th}}}{4R\left( P+1 \right)}\left[ 
  \theta -  
 \sqrt{{{\theta}^{2}}-8\left( P+1 \right)P{{{\tilde{n}}}_{0}}\left( {{{\tilde{n}}}_{0}}+1 \right)} \right],
\lb{Ne_st}
\eeq
where $\theta=2P{{{\tilde{n}}}_{0}}+\left( P+1 \right)\left( 1+{{{\tilde{n}}}_{0}} \right)+{{\beta }}$. We substitute the value of $N_e$ from Eq.~\rf{Ne_st} to Eq.~\rf{mean_n0} to find the mean photon number, $n_0$, for the carrier frequency $\omega_0$. Then,  from Eq.~\rf{n_pm1}, we can find the mean photon numbers, $n_{\pm 1}$,  for the carrier frequencies, $\omega_0\pm\omega_m$,
\beq
{{n}_{\pm 1}}={{\left[ \frac{2h{{\left( {{\omega }_{0}}\pm {{\omega }_{m}} \right)}^{2}}}{\omega _{c}^{2}-{{\left( {{\omega }_{0}}\pm {{\omega }_{m}} \right)}^{2}}} \right]}^{2}}\frac{{{n}_{0}}}{\left( 1\pm {{\omega }_{m}}/{{\omega }_{0}} \right)}\lb{n_pm1_mean}
\eeq
for each of the three  roots, $\omega_c^2$, of Eq.~\rf{char_eq}. Thus we found three modes of the laser with PTC,  each of which has a field of three carrier frequencies $\omega_0$, $\omega_0\pm\omega_m$. The $n_{0\omega}$ spectrum  is determined by Eq.~\rf{n0_sp_expl}, while $n_{\pm 1\omega}$ spectra 
 are given by Eq.~\rf{n_pm1}. 
 
  {Now, when the PTC laser field is found, we compare the PTC laser with a three-mode intra-cavity SFG/DFG. In SFG/DFG the cavity must support {\it three} modes of the same direction: the signal (S), the pump (P) and the generated (mix) mode with frequencies $\omega_{S,P}$ and $\omega_{mix} = \omega_P\pm\omega_S$, respectively. 
In a difference with SFG/DFG,  
a {\it single mode} with, in generally, many frequencies $\omega_0+ p\omega_m$, $p=0,\pm1, ...$ is generated in the PTC laser. 
We consider only frequencies $\omega_0\pm \omega_m$, and it makes our PTC laser similar with SFG/DFG. However the cavity must support at least three unidirectional modes in SFG/DFG, while we use {\it one cavity mode}, of the frequency $\omega_c$ in the PTC laser and generate a unidirectional field with many carrier frequencies, more than three, in generally. Our PTC laser cavity must support also the mode of frequency $\omega_m$, but in the direction perpendicular to the generation mode.

The possibility to generate many frequencies at the same spatial configuration of the field makes PTC laser, possibly, more flexible for some applications, then SFG/DFG. It may be in cases, when the unidirectional, focused field with many frequencies is required as, for example, for cascaded excitation of atoms or molecules. 

Another difference between PTC laser and SFG/DFG is that the energy comes from the active medium in the PTC laser, while it comes from the external (the pump) field in SFG/DFG. Specific choice of PTC laser or SFG/DFG scheme for multi-frequency generation depends on the application. PTC laser makes the choice of the multi-frequency generation schemes wider.}
 
 In the next section we will  consider the relationships between fields with different carrier frequencies within each mode. 
\section{\label{Sec3}Parameters of laser with PTC}
Let us estimate the parameters necessary for the effective modulation of the dielectric medium for  operating the laser with PTC. Suppose, that the nonlinear medium is  {like} a lithium niobate ($LiNbO_3$) with the second order dielectric susceptibility $\chi_2=200$~pm/V \ct{Zu2022}. In order to achieve effective modulation of PTC,  it is necessary to have $\delta\chi \sim \chi_2E_m \sim 1$. Therefore, the electric field strength must be $E_m\sim 1/\chi_2 = 5\cdot 10^9$~V/m, supposing the PTC modulation field  wavelength  {$\tilde{\lambda}_m$ is of the order of few microns, say, $\tilde{\lambda}_m=5$~mkm.} For a small cavity of the volume  {$\sim (\tilde{\lambda}_m/2)^3$ with the quality factor $Q_m=1000$}, such a field strength is attained for an input field of $4.7$~kWt power, which is  less than the power required for the second harmonic generation \ct{Evg_Dianov}. The intensity of the field  used for the PTC modulation can be approximately 19~MWt/cm$^2$ at the entrance of a small vertical cavity of the size  $\sim\tilde{\lambda}_m/2$ shown in Figure~\ref{Fig1}. 

\begin{figure}[thb]\bc
\centering
\includegraphics[width=12cm]{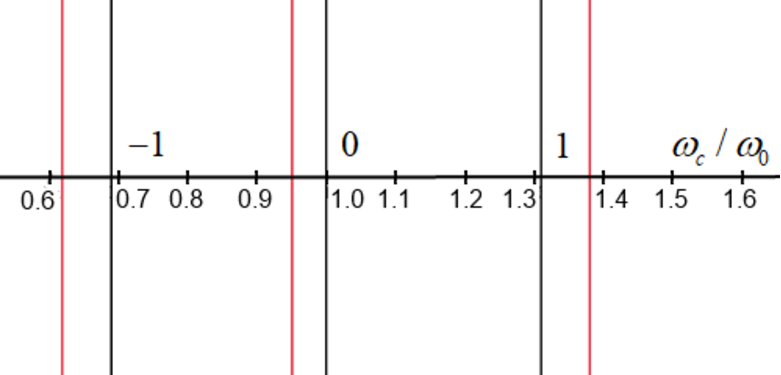}
\caption{ {  {HM} resonant frequencies $\omega_c/\omega_0$} of   {HM} $0$, $\pm 1$ 
are marked by vertical lines. Black lines denote $\omega_c$ without PTC (at $h=0$) red lines mark $\omega_c$ 
with PTC when $h> 0$ and a modulation of the nonlinear medium is on frequency $\omega_m$.  
}
\label{Fig2}\ec
\end{figure}

 {Above} it was determined  that three resonant radiation modes are possible in the laser with PTC, provided that the normalized second-order susceptibility coefficient   {$h=\pi\delta\chi$  is in the region $0<8{{h}^{2}}<1$, see the definition of $h$ before Eq.~\rf{char_eq_norm} and estimations of the $h$ value after Eq.~\rf{C_est}.} Each radiation mode has a specific resonant frequency $\omega_{c}$. The spectrum of each mode comprises three narrow maxima at carrier frequencies ${{\omega }_{0}}$, ${{\omega }_{0}}\pm {{\omega }_{m}}$. We denote  {the mode with the minimum value of $\omega_c$  by the index $-1$, the mode with the maximum $\omega_c$ by the index $1$, and the remaining mode -- by the index $0$, as shown in  Figure~\ref{Fig2}. Correspondingly, we will denote three roots of the equation \rf{char_eq} as $\omega_{c,p}^2$, $p=0,\pm 1$}.    {In case of  $8{{h}^{2}}>1$ the mode $-1$ has imaginary wave number, such a solution of Eq.~\rf{char_eq_norm} does not yield a real cavity resonance,  
 so this mode is not excited.} Modes $0$, $1$ are excited even at $8{{h}^{2}}>1$.

In the absence of a nonlinear medium, i.e. when $h=0$, we have $\omega _{c}^{2}=\omega _{0}^{2}$. In this  {case}, the mode 0 corresponds to the radiation of a conventional laser with a carrier frequency $\omega_0$, whilst the radiation in modes $\pm 1$ is absent.  
In the presence of a nonlinear medium that is modulated at the frequency $\omega_m$, the fields of carrier frequencies ${{\omega }_{0}}$, ${{\omega }_{0}}\pm {{\omega }_{m}}$ interact through the nonlinear medium. Consequently, radiation exists in each of the modes $0$, $\pm 1$. 

We use typical values for the parameters of quantum dot lasers with photonic crystal cavities \ct{FITSIOS201897}, when all three modes are possible in the PTC laser.  {Suppose that the field with the carrier frequency $\omega_0$, resonant with the active medium, has the wavelength $1.55$~mkm in the medium with the linear refractive index   {$n_r=3.42$}. We consider that the nonlinear medium    {interacting with VM} is modulated by the field of the wavelength $\tilde{\lambda}_m=5$~mkm. The size of   {VM} is about $\sim\tilde{\lambda}_m/2$. The size of   {HM} must be chosen specifically for each mode $p=0,\pm 1$, the size of   {HM} is about   $\tilde{\lambda}_{c,p}/2$, where $\tilde{\lambda}_{c,p} = 2\pi c/(n_r\omega_{c,p})$. For simplicity, we assume that   {HM}  in Figure~\ref{Fig1} is of the same size in all directions and that  $R_Z(z)R_{XY}(x,y) = \sin{\left( {2\pi z}/{{\tilde{\lambda }_{c,p}}} \right)}\sin{ \left( {2\pi x}/{{\tilde{\lambda }_{c,p}}} \right)}\sin{ \left( {2\pi y}/{{\tilde{\lambda }_{c,p}}} \right)}$ in Eq.~\rf{El_F_1}. The same we assume for   {VM}, so $R_{mZ} (z)R_{mXY}(x,y) = \sin{\left( {2\pi z}/{{\tilde{\lambda }_m}} \right)}\sin{ \left( {2\pi x}/{{\tilde{\lambda }_m}} \right)}\sin{ \left( {2\pi y}/{{\tilde{\lambda }_m}} \right)}$ in Eq.~\rf{VR_field}.
In order to estimate the coefficient \rf{coef_C} we approximate  $\tilde{\lambda}_{c,p}=\tilde{\lambda}_{c} = 1.55$~mkm. 
Then, the coefficient \rf{coef_C}, describing the overlap of the   {HM} and   {VM} can be estimated as}
\beq
  {C}\sim{{\left[ \frac{2}{{\tilde{\lambda }_c}}\int\limits_{0}^{{\tilde{\lambda }_c}/2}{\sin \left( \frac{2\pi z}{{\tilde{\lambda }_c}} \right)\sin \left( \frac{2\pi z}{{\tilde{\lambda }_{m}}} \right)dz} \right]}^{3}}=0.025.\lb{C_est}
\eeq
We take $\chi_2=200$~pm/V, $E_m=1/\chi_2 = 5\cdot 10^9$~V/m then $\delta\chi =1$ and $h=\pi \delta \chi {  {C}}=0.155<1/\sqrt{8}=0.354$, so all three   {HM} $0,\pm 1$ exist. We set the quality factor to $Q_m=10^3$ for all cavities and assume  that the cavity field decay rate  is $\kappa ={{\omega }_{0}}/2Q_m=6.1\cdot {{10}^{11}}$~rad/s, the same for all cavities and fields. We take the following values: the linewidth of the  active medium transitions $\gamma_{\perp} = 10^{12}$~rad/s; the upper active medium state decay rate $\gamma_{\parallel} = 10^9$~rad/s; the number of active emitters $N_0=100$; the semiclassical threshold population inversion $N_{th}=50$; and the linear refractive index of the medium in   {HM} and   {VM $n_r=3.42$}.
%
%%%%%%%%%%%%%%%%%%%%%%%%%%%%%%%%%%%%%%%%%%%%%%%
%
%
%%%%%%%%%%%%%%%%%%%%%%%%%%%%%%%%%%%%%%%%%%%%%%%

Figure~\ref{Fig2} allows us to compare the normalized  $\omega_c/\omega_0$ of the cavity modes $0,\pm 1$, which are solutions of the equation \rf{char_eq_norm}, without PTC, for $h=0$ (black vertical lines), and  with PTC for $h=0.155$ and the modulation of the nonlinear dielectric medium  (red lines). We can see that PTC noticeably    changes the   {HM} resonant frequencies $\omega_c$.   

\begin{figure}[thb]\bc
\centering
\includegraphics[width=7.5cm]{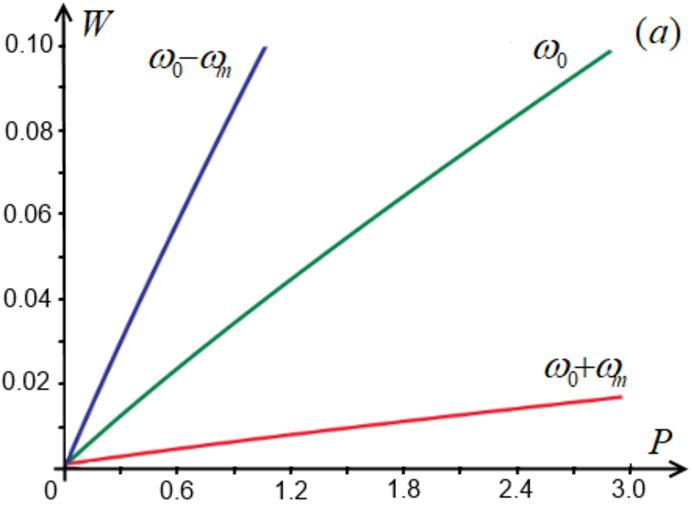}\hspace{1cm}\includegraphics[width=7.5cm]{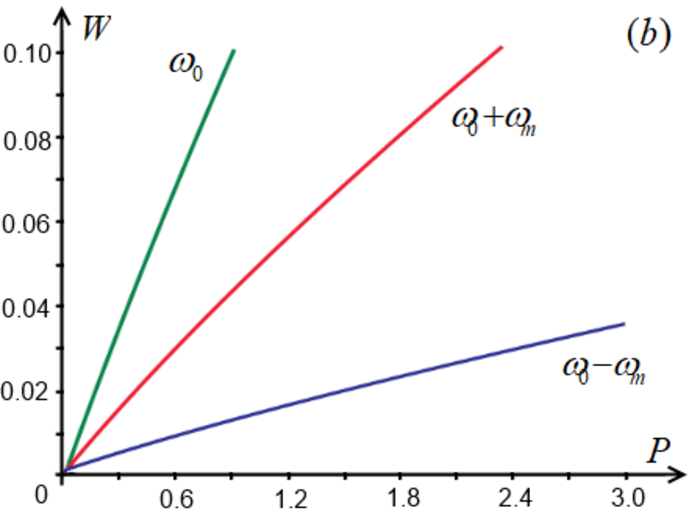}\\
\includegraphics[width=7.5cm]{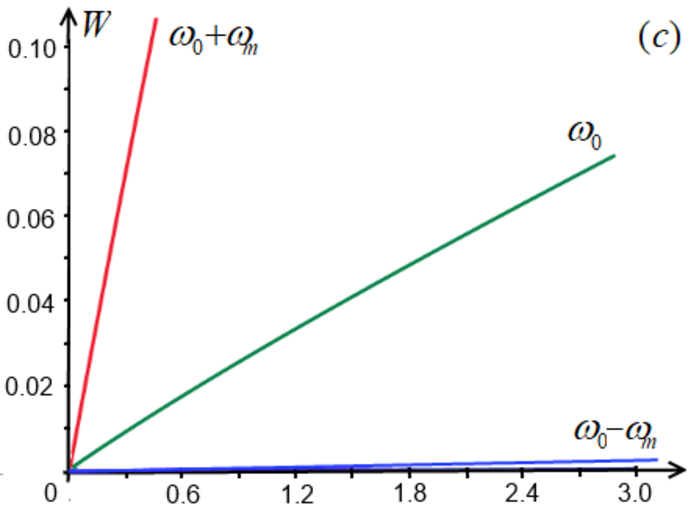}\hspace{1cm}\includegraphics[width=7.5cm]{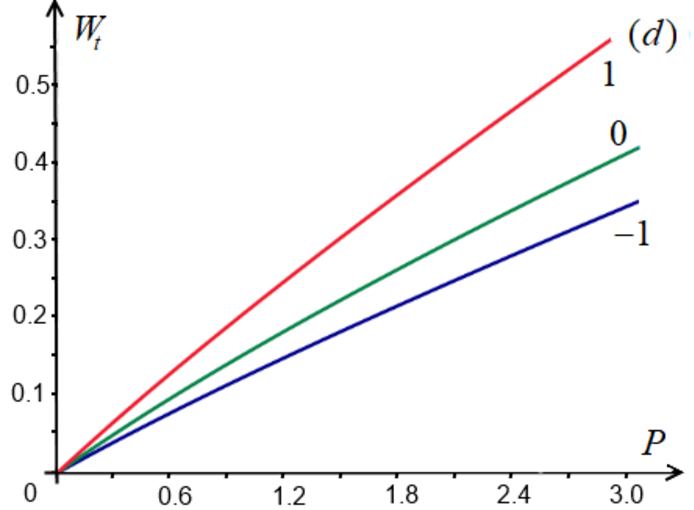}
\caption{The energies $W$ (in $\hbar\omega_0$ units) of fields with carrier frequencies $\omega_0$, $\omega_0\pm\omega_m$ (shown near curves) in   {HM}  $-1$ (a), $0$ (b) and $1$ (c) and the total energy $W_t$ of fields of these modes (shown near curves) (d) depended on the dimensionless rate $P$ of the pump of the upper states of the active medium.}
\label{Fig3}\ec
\end{figure}

The pictures in Figs~\ref{Fig3}~a-c show the average energies $W$ in units of $\hbar\omega_0$  for different types of the cavity fields  (green lines, red lines and blue lines).   {The fields have carrier frequencies (and $W$) of $\omega_0$ (and $W=n_0$), $\omega_0+\omega_m$ ($W=(1+\omega_m/\omega_0)n_1$), and $\omega_0-\omega_m$ ($W=(1-\omega_m/\omega_0)n_{-1}$) respectively.} The  pump rate $P$  of the active medium upper states is in the units of $\gamma_{\parallel}$. The pictures are for modes $-1$ (a), $0$ (b) and $1$ (c). Fig.~\ref{Fig3}~d shows the total energy   {$W_t(P)=n_0+(1+\omega_m/\omega_0)n_1+(1-\omega_m/\omega_0)n_{-1}$} (in $\hbar\omega_0$ units) of the field in   {HM}  $0$ (green line), $1$ (red) and $-1$ (blue).  
%
%%%%%%%%%%%%%%%%%%%%%%%%%%%%%%%%%%%%%%%%%%%%%%%
%
%
%%%%%%%%%%%%%%%%%%%%%%%%%%%%%%%%%%%%%%%%%%%%%%%
As it is seen in Fig.~\ref{Fig3}, the laser with PTC is working in a quantum regime, with a mean photon number  {in   {HM}} is less than 1 for each carrier frequency.   {At such a small photon number in the cavity the PTC laser works in the linear regime.}  Modes $-1$, $0$ and $1$  {radiate} best at frequencies $\omega_0-\omega_m$, $\omega_0$ and 
$\omega_0+\omega_m$, respectively. The efficiency of the radiation at all carrier frequencies is close to each other in each mode, see Fig.~\ref{Fig3}d. The field with frequency $\omega_0-\omega_m$  in mode $1$ has much less energy ($W\sim10^{-4}$ in Fig.~\ref{Fig3}c) than the fields with other carrier frequencies. This is because of the quantum  $\hbar(\omega_0-\omega_m)$ is the smallest one and the frequency $\omega_0-\omega_m$ is very far from the frequency $\omega_0+\omega_m$ emitted most effectively in the mode $1$, see Fig.~\ref{Fig3}c. The total radiation energy $W_t$ in Fig.~\ref{Fig3}~d is roughly the same for each mode,  the energy  distribution across fields with different carrier frequencies changes from mode to mode.

\begin{figure}[thb]\bc
\centering
\includegraphics[width=7.5cm]{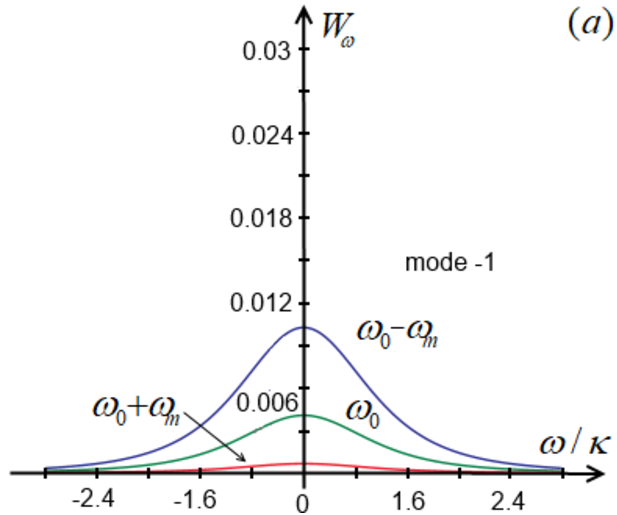}\hspace{1cm}\includegraphics[width=7.5cm]{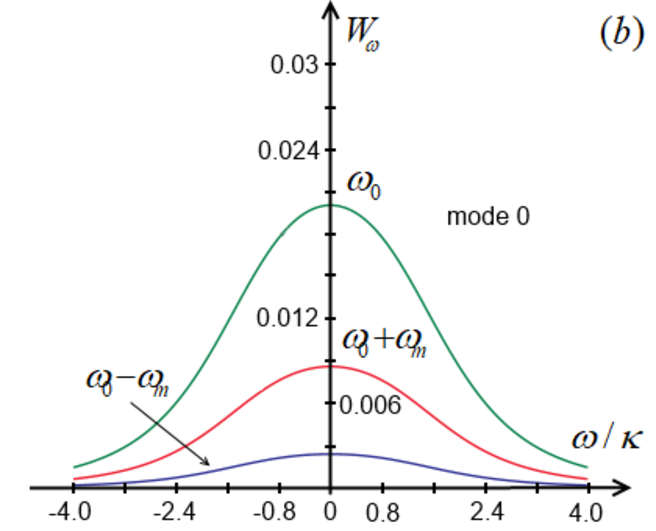}\\
\includegraphics[width=7.5cm]{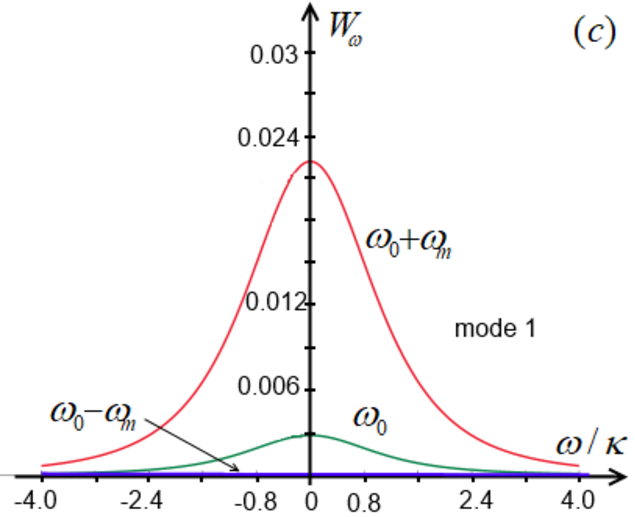}
\caption{The spectra $W_{\omega}$ (in $\hbar\omega_0$ units) of the fields in   {HM}, centered at the carrier frequencies (written near curves) are shown in figures $a$, $b$, and $c$, for   {HM} $-1$, $0$, and $1$,  respectively. $\omega$ is the detuning from each of carrier frequency. }
\label{Fig4}\ec
\end{figure}

Let us consider lasing spectra and estimate  lasing line widths.   {Field spectra 
$W_{\omega} = n_{0\omega}$ for the carrier frequency $\omega_0$ and $W_{\omega} = (1\pm\omega_m/\omega_0)n_{\pm 1\omega}$ for the carrier frequency $\omega_0\pm\omega_m$ are shown in Figures~\ref{Fig4}~a-c for  {the pump} $P=1$; $n_{\pm 1\omega}$, $n_{0\omega}$ are given by expressions \rf{n_pm1} and \rf{n0_sp_expl}, respectively.} For convenience, we align the central (carrier) frequencies of each spectrum.  The spectra are functions of  the detuning, $\omega$, from the carrier frequencies. The detunings are normalized by the decay rate, $\kappa$, of the cavity field. The blue, green and red curves correspond to the carrier frequencies $\omega_0-\omega_m$, $\omega_0$ and  $\omega_0+\omega_m$. As can be seen in Fig.~\ref{Fig4}, the linewidths of all the spectra are of the order of $\kappa\ll\omega_0$, so each spectrum is relatively narrow.   The narrow  line width of the spectra with $\omega_0\pm\omega_m$ carrier frequencies is explained by the result \rf{n_pm1}, which shows that the spectra $n_{{\pm 1}\omega}$ are proportional to a narrow $n_{0\omega}$ lasing spectrum.
%
%%%%%%%%%%%%%%%%%%%%%%%%%%%%%%%%%%%%%%%%%%%%%%%
%
%
%%%%%%%%%%%%%%%%%%%%%%%%%%%%%%%%%%%%%%%%%%%%%%%

\begin{figure}[thb]\bc
\centering
\includegraphics[width=7.5cm]{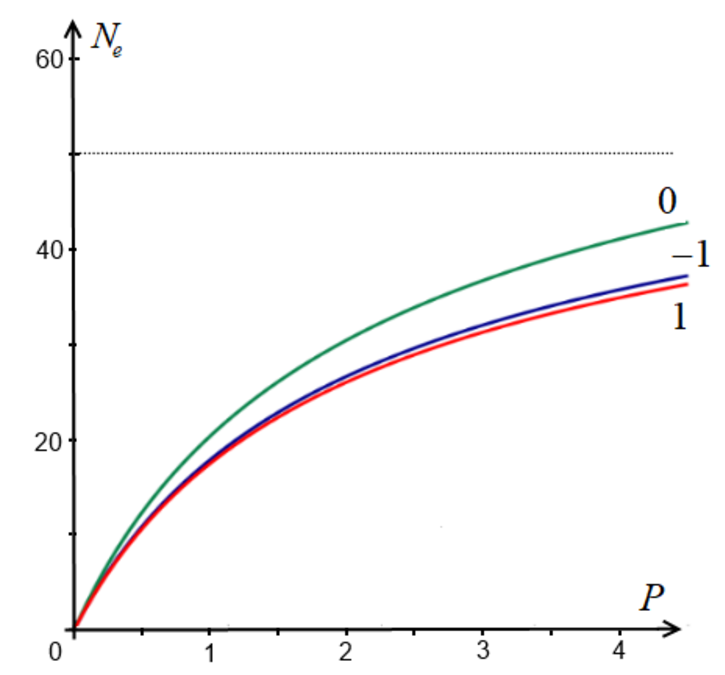}
\caption{Population $N_e$ of the upper states of two-level emitters depended on the pump $P$ for modes with indexes $-1$, $0$, and $1$, shown near curves. Smaller $N_e$ means that more energy comes to the emitted field and the lasing is more efficient. The  population inversion exists for $N_e$ values above the horizontal line.  }
\label{Fig5}\ec
\end{figure}

Fig.~\ref{Fig5} shows the dependence of the population $N_e$ of the upper states of the active medium on the dimensionless pump $P$. We  see that $N_e<N_0/2=50$. This means that the population inversion is not reached at the pump rates considered here. Therefore, the laser with PTC operates as a light emitted diode. The minimum value of $N_e$ is reached, when the maximum energy is taken by fields from the active medium, and the lasing is most efficient. The minimum $N_e$ and the maximum lasing efficiency are reached for the mode $1$, which has the most energy in the field  of the largest carrier frequency $\omega_0+\omega_m$, as shown also in Figs.~\ref{Fig3}c and \ref{Fig4}c. 
%
%%%%%%%%%%%%%%%%%%%%%%%%%%%%%%%%%%%%%%%%%%%%%%%
%
%
%%%%%%%%%%%%%%%%%%%%%%%%%%%%%%%%%%%%%%%%%%%%%%%
%
\section{\label{Sec4}Discussion of results}
The key difference between the proposed laser with PTC and the usual laser is in the spectrum of the lasing mode. The laser with PTC emits fields with three different carrier frequencies: $\omega_0$ and $\omega_0\pm\omega_m$ in the lasing mode,  {while the usual laser without PTC has only one carrier frequency in the mode}. Here, $\omega_0$ is the transition frequency of the active medium and $\omega_m$ is the modulation frequency of PTC.
We generalise the standard approach for calculating the mode  {resonant frequency} 
for the usual laser to the laser with PTC.     {This resonant frequency} is a solution of the  equation \rf{char_eq}. We identified three lasing modes, which are indicated by the indexes   {$p=0,\pm 1$}. There are three  {emitting} frequencies in each of these modes. In the absence of PTC, modes $0$ and $\pm 1$ are the independent cavity modes, with the  {resonant} frequencies of $\omega_0$, $\omega_0\pm\omega_m$, respectively. When PTC is present and  modulated by an external field, the interaction of fields with different carrier frequencies in the nonlinear medium generates all these three fields in each mode. In each mode, the field with the carrier frequency, which exist without PTC, is generated more efficiently than the fields of the other frequencies.  {So the field of frequency $\omega_0-\omega_m$ is generated most effectively in the mode $p=-1$, the field of frequency $\omega_0$ -- in the mode $p=0$ and  the field of frequency $\omega_0+\omega_m$ -- in the mode $p=1$, see  Figs.~\ref{Fig3}~a-c.} The width of the peaks near each carrier frequency is small, on the order of the cavity decay rate $\kappa$. This is much smaller than both $\omega_0$ and  $\omega_m$. The total energy emitted is approximately the same for all modes (see Fig.~\ref{Fig3}~c).

A laser with a PTC that generates a field with three well-separated peaks  {in the field spectrum} could be useful for spectroscopy and resonant photoionisation. The distance between the peaks can be altered by adjusting the modulation frequency,  $\omega_m$. The mode with the maximum energy at a given frequency can be selected according to the intended application.  For example, mode $-1$, which has the maximum energy in the field at the lowest carrier frequency $\omega-\omega_m$, is suitable for resonantly enhanced multi-photon ionisation (REMPI) via the intermediate, short-lived excited state \ct{Kuraishi2013}.

In principle, more than three carrier frequencies can be generated in the laser with PTC as set out in the general theory of PTC \ct{Asgari:24}.

A laser with a PTC and a small cavity operates in the quantum regime, where the average number of photons for each carrier frequency in each mode is less than one. This is why we use here the quantum theory of  micro-lasers, described in \ct{Andre:19,Protsenko_2021,PhysRevA.105.053713}.  In the current step of research of the laser with PTC, we neglect active medium population fluctuations. This paper's approach lets us explicitly determine the mean photon numbers and the lasing field spectra. We will take population fluctuations into account, following \ct{Protsenko_2021}, and, in the future, investigate collective Rabi splitting in a laser with PTC, following \ct{Andre:19}.

We estimate that the effective generation of modes with three carrier frequencies is possible under realistic conditions in a photonic crystal laser with a quantum dot active medium and PTC with a semiconductor nonlinear medium. Effective modulation of the refractive index in a nonlinear medium with second-order dielectric susceptibility is possible with a modulating field power of around 5~kWt and a field intensity of approximately 20 MWt/cm$^2$ at the entrance of the PTC cavity. Such power and intensity are practically possible; they are smaller than the power and intensity of the field in second-harmonic generation experiments \ct{Evg_Dianov}.  

It will be interesting to combine the laser with PTC and the radiation source for PTC modulation in the same miniature setup. In order to achieve this, the active medium with the resonant frequency $\omega_m$ must be added to the vertical cavity.   A modulating field with sufficiently high power density can be realized in a small-volume vertical cavity.  In the future, we will analyze such compact PTC micro-laser scheme combined with the  PTC modulator.
\section{\label{Sec5}Conclusion}
 {We proposed a photonic time crystal microlaser consisting of two-level active oscillators (e.g., quantum dots) embedded in a nonlinear medium with second-order dielectric susceptibility. This active medium is located within a small photonic crystal cavity at the intersection of horizontal and vertical one-dimensional photonic crystals. The small cavity provides an electromagnetic field with sufficient energy density to effectively modulate the PTC under realistic experimental conditions. 
We used a quantum optical approach to analyze this PTC laser. Specifically, we identified the laser's field modes with PTC, its lasing field spectra, and the mean photon numbers for each carrier frequency in these modes. We found that this laser produces a field spectrum with narrow peaks at three distinct carrier frequencies.  
A multi-frequency laser with PTC has a variety of potential applications, including spectroscopy and multi-step photoionization. We hope this theoretical consideration encourages experimental research into microlasers with PTC.}
 
\medskip

\bibliography{myrefs}

\end{document}